\documentclass[aps,pra,twocolumn,amsmath,amssymb,superscriptaddress,showpacs]{revtex4-1}
\usepackage{graphics}
\usepackage{graphicx}
\usepackage{bbm}
\usepackage{amsfonts}
\usepackage{amssymb}
\usepackage{epsfig}
\usepackage{xcolor}

\newcommand{\ket}[1]{|#1\rangle}

\begin{document}

\title{Quantum coherence in a compass chain under an alternating
       magnetic field}

\author {Wen-Long You}
\affiliation{\mbox{College of Physics, Optoelectronics and Energy,
Soochow University, Suzhou, Jiangsu 215006, China}}
\affiliation{Jiangsu Key Laboratory of Thin Films, Soochow University,
Suzhou 215006, China}

\author{Yimin Wang}
\affiliation{Communications Engineering College,
Army Engineering University, Nanjing, Jiangsu 210007, China}

\author {Tian-Cheng Yi}
\affiliation{\mbox{College of Physics, Optoelectronics and Energy,
Soochow University, Suzhou, Jiangsu 215006, China}}

\author {Chengjie Zhang}
\affiliation{\mbox{College of Physics, Optoelectronics and Energy,
Soochow University, Suzhou, Jiangsu 215006, China}}

\author{Andrzej M. Ole\'s}
\affiliation{Max Planck Institute for Solid State Research,
             Heisenbergstrasse 1, D-70569 Stuttgart, Germany}
\affiliation{\mbox{Marian Smoluchowski Institute of Physics, Jagiellonian
             University, Prof. S. \L{}ojasiewicza 11, PL-30348 Krak\'ow, Poland}}

\date{25 April, 2018}

\begin{abstract}
We investigate quantum phase transitions and quantum coherence in
a quantum compass chain under an alternating transverse magnetic
field. The model can be analytically solved by the Jordan-Wigner
transformation and this solution shows that it is equivalent to
a two-component one-dimensional (1D) Fermi gas on a lattice.
We explore mutual effects of the staggered magnetic interaction and
multi-site interactions on the energy spectra and analyze the ground
state phase diagram. We use quantum
coherence measures to identify the quantum phase transitions. Our
results show that $l_1$ norm of coherence fails to detect faithfully
the quantum critical points separating a gapped phase from a gapless
phase, which can be pinpointed exactly by relative entropy of
coherence. Jensen-Shannon divergence is somewhat obscure at
exception points. We also propose an experimental realization of
such a 1D system using superconducting quantum circuits.
\end{abstract}


\maketitle

\section{Introduction}

Many physical phenomena in quantum information science have evolved
from being of purely theoretical interest to enjoying a variety of
uses as resources in quantum information processing tasks. Throughout
the development of the resource theory of entanglement, various
measures were established. However, entanglement is not a unique
measure of quantum correlation because separable states can have
nonclassical correlations. The concept of quantum coherence has
recently seen a surge of popularity since it serves as a resource
in quantum information tasks \cite{Streltsov2017}, similar to other
well-studied quantum resource such as the entanglement
\cite{Osterloh2002}, quantum correlations \cite{Ollivier01}, and
the randomness \cite{Bera17}.
Baumgratz \textit{et al.} \cite{Baumgratzq14} introduced a rigorous
framework for the quantification of coherence based on resource
theory and identified easily computable measures of coherence.
Quantum coherence resulting from quantum state superposition plays
a key role in quantum physics, quantum information processing, and
quantum biology.

Ideally the coherence of a given state is measured as its distance to
the closest incoherent state \cite{Girolami14,Bromley15,Streltsov15,Chitambar16,Killoran16,Winter16,
Napoli16,Bagan16,Ma16,Streltsov16,Chitambar16a,Chitambar16b,Liu17a,
Bu17,Theurer17,Wang17,Zhang17,Hu17}.
Coherence properties of a quantum state
are usually attributed to the off-diagonal elements of its density
matrix with respect to a selected reference basis. Among a few popular
measures, there are three recently introduced coherence measures,
namely, relative entropy of coherence, $l_1$ norm quantum coherence
\cite{Baumgratzq14}, and Jensen-Shannon (JS) divergence \cite{Rad16}.
The JS divergence and the $l_1$ norm of coherence obey the symmetry
axiom of a distance measure while the relative entropy does not obey
such distance properties. The $l_1$ norm of coherence sums up
absolute values of all off-diagonal elements $\rho_{i,j}$
(with $i\neq j$) of the density matrix $\rho$, that is,
\begin{eqnarray}
C_{l_1} (\rho)&=&\sum_{i\neq j} \vert \rho_{i,j}\vert,
\label{eq:measure1}
\end{eqnarray}
where $\rho_{i,j}$ is the element of the density matrix with $i$ and
$j$ being the row and the column index. $C_{l_1} (\rho)$ is a geometric
measure that can be used as a formal distance measure. In parallel,
the relative entropy has been established as a valid measure
of coherence for a given basis:
\begin{eqnarray}
C_{\rm re}({\rho}) &=&S({\rho}\vert\vert{\rho}_{\rm diag})
= S({\rho}_{\rm diag}) - S({\rho}),
\label{eq:measure2}
\end{eqnarray}
where $S(\rho)=-{\rm Tr} (\rho\log_2\rho)$ stands for the von Neumann
entropy and ${\rho}_{diag}$ is the incoherent state obtained from
${\rho}$ by removing all its off-diagonal entries. $C_{l_1}$ and
$C_{\rm re}$ are known to obey strong monotonicity for all states.
Since $C_{\rm re}$ is similar with relative entropy of entanglement,
it has a clear operational interruption as the distillable coherence.
Meanwhile, $C_{l_1}$ takes an operational interpretation as the maximum
distillable coherence from a resource theoretical viewpoint
\cite{Rana17,Zhu17}. It was shown that $C_{l_1}$ is an upper bound for
$C_{\rm re}$ for all pure states and qubit states \cite{Rana16}.
Moreover, the measure of quantum coherence based on the square root of
the JS divergence is given by
\begin{equation}
C_{\rm JS}({\rho})=
\sqrt{S\left(\frac{\rho_{\rm diag}+{\rho}}{2}\right)
-\frac12 S({\rho}_{\rm diag}) -\frac12 S({\rho})}.
\label{eq:measure3}
\end{equation}
The JS divergence is known to be a symmetric and bounded distance
measure between mixed quantum states \cite{Majtey05} and is exploited
to study shareability of coherence \cite{Rad16}. We remark that these
coherence measures are all basis dependent \cite{Malvezzi16}.

Currently, approaches adopted from quantum information theory are being
tested to explore many-body theory from another perspective and vice
versa. Various quantum resource measures have been exploited to
characterize the state of many-body systems and the associated quantum
phase transitions (QPTs). A QPT occurs at zero temperature and is
engraved by a qualitative change solely due to quantum fluctuations as
a non-thermal parameter is varied. Quantum entanglement and quantum
discord have been proven to be fruitful to investigate QPTs. For
instance, entanglement entropy changes at some (but not all) QPTs
\cite{You15}. Investigation of entanglement spectra is therefore
very useful and helps to identify a possible QPT when the entanglement
entropy in the ground-state changes by a finite value when Hamiltonian
parameters are varied.
One of easily accessible parameters is a magnetic field which might
control producing quantum matter near a quantum critical point (QCP)
in spin chains \cite{Bla18}. Small systems of interacting spins in a
two-dimensional (2D) compass model with perturbing interactions could
also be used for quantum computation~\cite{Tro12}.

Quantum coherence has emerged from an information physics perspective
to address different aspects of quantum correlation in a many-body
system. Comparing with entanglement measures, quantum coherence is
expected to be capable of detecting QPTs even when the entanglement
measures fail to do so. One can easily recognize that entanglement
may be a form of coherence and the converse is not necessarily valid.
For instance, a product state
($\vert 0\rangle+\vert 1\rangle)\otimes(\vert 0\rangle+\vert 1\rangle$),
for a two-qubit system carries coherence but not entanglement. That is
to say, quantum coherence incarnates a different feature of a quantum
state from entanglement. On the other hand, coherence measures can be
used as a resource in quantum computing protocols, and one may claim
that they are more fundamental. So a comparative study of these
measures for characterizing QPTs in various spin chain models is a
potential research topic, which may be valuable in both physical theory
and experiment.

To test the validity of this approach, the Ising model is the most
transparent example of the importance of exactly soluble models as
guides along this difficult path. All coherence measures are able to
locate the Ising-type second-order transition. Here we consider another
prominent model dubbed as a one-dimensional (1D) compass model for a
$p$-wave superconducting chain, which sustains more complex physical
phenomena than the Ising model, such as macroscopic degeneracy
\cite{Brzezicki07}, pure classical features \cite{You12}, and
suppressed critical revival structure~\cite{Jafari17}.

The purpose of this paper is to investigate QPTs and quantum coherence
in the 1D quantum compass model (QCM) under an alternating transverse
magnetic field. The motivation is twofold. On the one hand, previous
investigations revealed that the 1D compass model could exhibit
miscellaneous phases via modulation of external fields. An exotic
spin-liquid phase can emerge through a Berezinskii-Kosterlitz-Thouless
(BKT) QPT under a uniform magnetic field. We would like to verify
whether such a transition is robust under realistic inhomogeneity of
external fields~\cite{You14}. The calculations take into account both
uniform and staggered fields. On the other hand, its exact solvability
provides a suitable testing ground for calculating accurately coherence
measures to detect QPTs. We investigate the coherence of this model
in the thermodynamic limit and its connection to QPTs.

The remaining of the paper is structured as follows. An overview of the
1D compass model with staggered magnetic fields is presented in Sec.
\ref{sec:model}. We consider the cases in the presence of a uniform
and an alternating transverse magnetic field, and discuss a possible
experimental realization using superconducting quantum circuits in Sec.
\ref{sec:exp}. The model is extended by adding three-site interactions
in Sec. \ref{sec:3site}. Next the model is exactly solved and QPTs are
studied. We present the calculations of quantum coherence measures in
Sec. \ref{sec:measures}. A final discussion and summary are presented
in Sec. \ref{sec:con}.

\section{The Model and Its Analytical Solution}
\label{sec:model}

We begin with a generic 1D QCM \cite{Brzezicki07} on a ring of $N$
sites, where $N$ is even. The Hamiltonian describes a competition
between two pseudospin $\tau=1/2$ components,
$\{\sigma^x_i,\sigma^y_i\}$, which reads,
\begin{eqnarray}
\cal{H}_{\rm QCM}&=& \sum_{i=1}^{N/2}
\left(J_{1}\,X_{2i-1,2i}+J_{2}\,Y_{2i,2i+1}\right),
\label{Hamiltonian1}
\end{eqnarray}
and has the highest possible frustration of interactions.
Here $X_{i,j}\equiv\sigma_i^x\sigma_j^x$,
     $Y_{i,j}\equiv\sigma_i^y\sigma_j^y$, and $\sigma_i^{\alpha}$
is a Pauli matrix. $J_1$ ($J_2$) stands for
the amplitude of the nearest neighbor interaction on odd (even) bonds.
This model owns a particular intermediate symmetry, which allows for
$N/2$ mutually commuting $\mathbb{Z}_2$ invariants $Y_{2i-1,2i}$
($X_{2i,2i+1}$) in the absence of the transverse field term,
and presents distinct features. The ground state possesses a
macroscopic degeneracy of at least $2^{(N/2-1)}$ in the structure of
the spin Hilbert space \cite{Brzezicki07,You14}. The intermediate
symmetries also admit a dissipationless energy current \cite{Qiu16}.
The 1D QCM Eq. (\ref{Hamiltonian1}) can be
transformed to the fermion language and next diagonalized, see the
Appendix. In fact, the model can be described by a two-component 1D
Fermi gas on a lattice as displayed in Eq. (\ref{Hamiltonian2}).

The 1D QCM in Eq. (\ref{Hamiltonian1}) may be supplemented by a
possibly spatially inhomogeneous Zeeman field $\vec{h}_{i}$, given by
\begin{eqnarray}
\cal{H}_{\rm h}&=& \sum_{i=1}^{N}\vec{h}_{i}\cdot\vec{\sigma}_{i}.
\label{Hamiltonian3}
\end{eqnarray}
In a realistic structure, the crystal fields surrounding the
odd-indexed sites and even-indexed sites are different. The presence
of two crystallographically inequivalent sites on each chain with a low
symmetry of the crystal structure leads to staggered gyromagnetic
tensors.

It has been recently shown that a spatially varying magnetic
field can be induced by an effective spin-orbit interaction. The
alternating  spin  environment  is  represented  by  the staggered
Dzyaloshinskii-Moriya interaction and Zeeman terms.
The staggered magnetic field plays an important role in understanding
the field dependence of the gap in Cu benzoate antiferromagnetic chain
\cite{Affleck99,Fledderjohann96,Zhao03,Kenzelmann04} and in
Yb$_4$As$_3$ \cite{Kohgi01}. We remark that there is increasing
interest in the effects of the staggered field motivated by the
experimental work on a number of materials. The interplay between the
staggered Zeeman fields and dimerized hopping on the topological
properties of Su-Schrieffer-Heeger (SSH) model has received much
attention only recently \cite{SSH}, after the rapid progress in the
synthesis of 1D heterostructures \cite{Sau10,Alicea10,Das12,Perge14}.

Without the loss of  generality, we consider a staggered magnetic field,
i.e., $\vec{h}_{2i-1}$=$h_1\hat{z}$, $\vec{h}_{2i}$=$h_2\hat{z}$,
$\forall i=1,\cdots,N/2$. Be aware that $h_{i}=g_{i} \mu_B B_{i}$ here
is the reduced magnetic field containing the $g$-factor and the Bohr
magneton $\mu_{\rm B}$. An effective staggered magnetic field might be
attributed to the alternating $g$ tensor in an applied uniform field
\cite{Feyerherm2000}. Thereby, we define an average magnetic field
$h=(h_1+h_2)/2$ and a field difference $\delta=(h_2-h_1)/2$.
The 1D compass model in external field,
\begin{eqnarray}
\cal{H}&=&\cal{H}_{\rm QCM}+\cal{H}_{\rm h},
\label{com}
\end{eqnarray}
is exactly soluble and we obtain its zero-temperature phase diagram,
see below. For the sake of clarity, we briefly describe the procedure
to diagonalize the Hamiltonian Eq. (\ref{com}) exactly in the Appendix.

Thus the Hamiltonian (\ref{com}) in the
Bogoliubov-de Gennes (BdG) form in terms of Nambu spinors is:
\begin{eqnarray}
\cal{H} &=&  \frac{1}{2} \sum_{k}
\Upsilon_k^{\dagger}
\hat{H}_k
\Upsilon_k, \label{FT2}
\end{eqnarray}
where
$\Upsilon_k^{\dagger}=(a_k^{\dagger},b_k^{\dagger},a_{-k}^{},b_{-k}^{})$.
In this circumstance, the Hamiltonian (\ref{com}) reads
\begin{eqnarray}
 \hat{H}_k
&=& 2\delta
\Gamma_{zz} - 2 h  \Gamma_{z0} +T^r_k(\Gamma_{zx}-\Gamma_{yy})
-T^i_k(\Gamma_{zy}+ \Gamma_{yx}),\nonumber \\
\label{HamiltonianMatrix2}
\end{eqnarray}
with $\Gamma_{ab}=\tau^a \otimes \sigma^b$,  $\forall$ $a, b= x, y, z$,
and $\tau^{x,y,z}$/$\sigma^{x,y,z}$
being the Pauli matrices acting on particle-hole space and spin space,
respectively, and $\tau^{0}=\sigma^{0}$ is a 2$\times$2 unit matrix.
Here $T_k^r$ and $T_k^i$ are the real and imaginary parts of
$T_k=J_1+J_2e^{ik}$.
The BdG Hamiltonian Eq. (\ref{HamiltonianMatrix2}) respects a
particle-hole symmetry defined as
${\cal C}  \hat{H}(k) {\cal C}^{-1} = - \hat{H}(-k)$ with
${\cal C} = \Gamma_{x0} \cal{K}$, where $\cal{K}$ is the complex
conjugate operator. As a consequence the energy levels appear in
conjugate pairs such as $\varepsilon(k)$ and $-\varepsilon(-k)$.
The diagonal form of the Hamiltonian Eq. (\ref{HamiltonianMatrix2})
is then given by,
\begin{eqnarray}
{\cal H}=\sum_{k}\sum_{j=1}^{2}  \varepsilon_{k,j}
\left(\gamma_{k,j}^{\dagger}\gamma_{k,j}^{}-\frac{1}{2}\right).
\label{diagonalform}
\end{eqnarray}
The spectra consist of two branches of energies $\varepsilon_{k,j}$
(with $j=1,2$), given by the following expressions:
\begin{eqnarray}
\varepsilon_{k,1(2)}=
\sqrt{\vert T_k \vert^2+ 4 h^2}\pm \sqrt{\vert  T_k\vert^2 +4 \delta^2}.
\label{excitationspectrum2}
\end{eqnarray}
The ground-state energy per site for $h>\delta$ may be expressed as
\begin{eqnarray}
e_0 = - \frac{2}{N}\sum_{k} \sqrt{J_1^2+J_2^2+2J_1 J_2\cos k+ 4 h^2}.
\label{E0expression}
\end{eqnarray}
The advantage of the result given by Eq. (\ref{E0expression}) is that
$e_0$ is independent of $\delta$ as well as of the signs of $J_1$ and
$J_2$. The intersite correlators are given by the Hellmann-Feynman
theorem:
\begin{eqnarray}
\langle \sigma^x_{2i-1} \sigma^x_{2i}\rangle=-
\frac{2}{N}\sum_k\frac{J_1+J_2\cos k}{\sqrt{J_1^2+J_2^2+2J_1 J_2\cos k+4h^2}},
\nonumber \\
\langle \sigma^y_{2i} \sigma^y_{2i+1}\rangle=
-\frac{2}{N}\sum_k\frac{J_2+J_1\cos k}{\sqrt{J_1^2+J_2^2+2J_1 J_2\cos k+4h^2}}.
\nonumber\\\label{intersitecorrelators}
\end{eqnarray}
For $\delta>h$, an interchange between $h$ and $\delta$ is performed
in Eqs. (\ref{E0expression}-\ref{intersitecorrelators}).

Since a QPT occurs only when the gap closes, looking for gapless
points in the energy spectrum may indicate this transition. The lower
mode $\varepsilon_{k,2}$ reduces to a zero-energy flat band for
$h=\pm\delta$, corresponding to either $h_{2i-1}=0$ or $h_{2i}=0$.
This undermines the limited condition for the existence of a
macroscopic degeneracy in the ground-state manifolds. The zero-energy
flat band is fragile against an infinitesimal external uniform magnetic
field for $\delta=0$. A uniform field will remove the ground-state
degeneracy and the bands are no longer degenerate. The result here
implies that a magnetic field applied on one sublattice still makes
the zero-energy flat band intact.

Interestingly, the model possesses local symmetries that one can find
in the absence of field terms at odd sites. If field $h_{2i-1}$ is
vanishing at odd sites and at even sites it takes any random values,
then any eigenstate has $2^{(N/2-1)}$ degeneracy for a ring of length
$N$. These degeneracies follow from the symmetry operators,
\begin{equation}
S_{2i}\equiv\sigma^y_{2i-1}\otimes\sigma^z_{2i}\otimes\sigma^x_{2i+1},
\label{Si}
\end{equation}
and are activated when the field is absent at odd sites. Such
symmetry operators (\ref{Si}) anti-commute for the neighbors, i.e.,
$\{S_{2i},S_{2(i+1)}\}=0$, while they commute otherwise.

\section{Possible experimental realization
         using superconducting quantum circuits}
\label{sec:exp}

The unique features of this rich model (\ref{com}) motivate us to
consider its possible physical implementations to advance our
understandings. It is well known that superconducting circuit systems
have become one of the leading platforms for scalable quantum
computation, quantum simulation and demonstrating quantum optical
phenomena because of its exotic properties such as controllability,
flexibility, scalability, and compatibility with micro-fabrication
\cite{Wen05,you2005pt,Dev13}. Various models of many-body systems have
been proved to be able to be simulated by superconducting circuits,
such as the Kitaev lattice~\cite{You10}, the Heisenberg spin model
\cite{Her14}, the fermionic model \cite{Bar15}, the 1D Ising model
\cite{Bar16} and anisotropic quantum Rabi model \cite{Wang18}.
In our case, the 1D compass chain in Eq. (\ref{com}) can be built from
superconducting charge qubits, each of which is composed of a direct
current superconducting quantum interference device (dc SQUID) with two
identical Josephson junctions. For $i$th charge qubit, the gate voltage
$V_{gi}$ applied through the gate capacitance $C_{gi}$ can be used to
control the charge, and the magnetic flux $\Phi_{ei}$ piercing the
SQUID can be used to control the effective Josephson energy,
$E_{J}(\Phi_{ei})=2E_J\cos(\pi\Phi_{ei}/\Phi_0)$.

\begin{figure}
\includegraphics[width=\columnwidth]{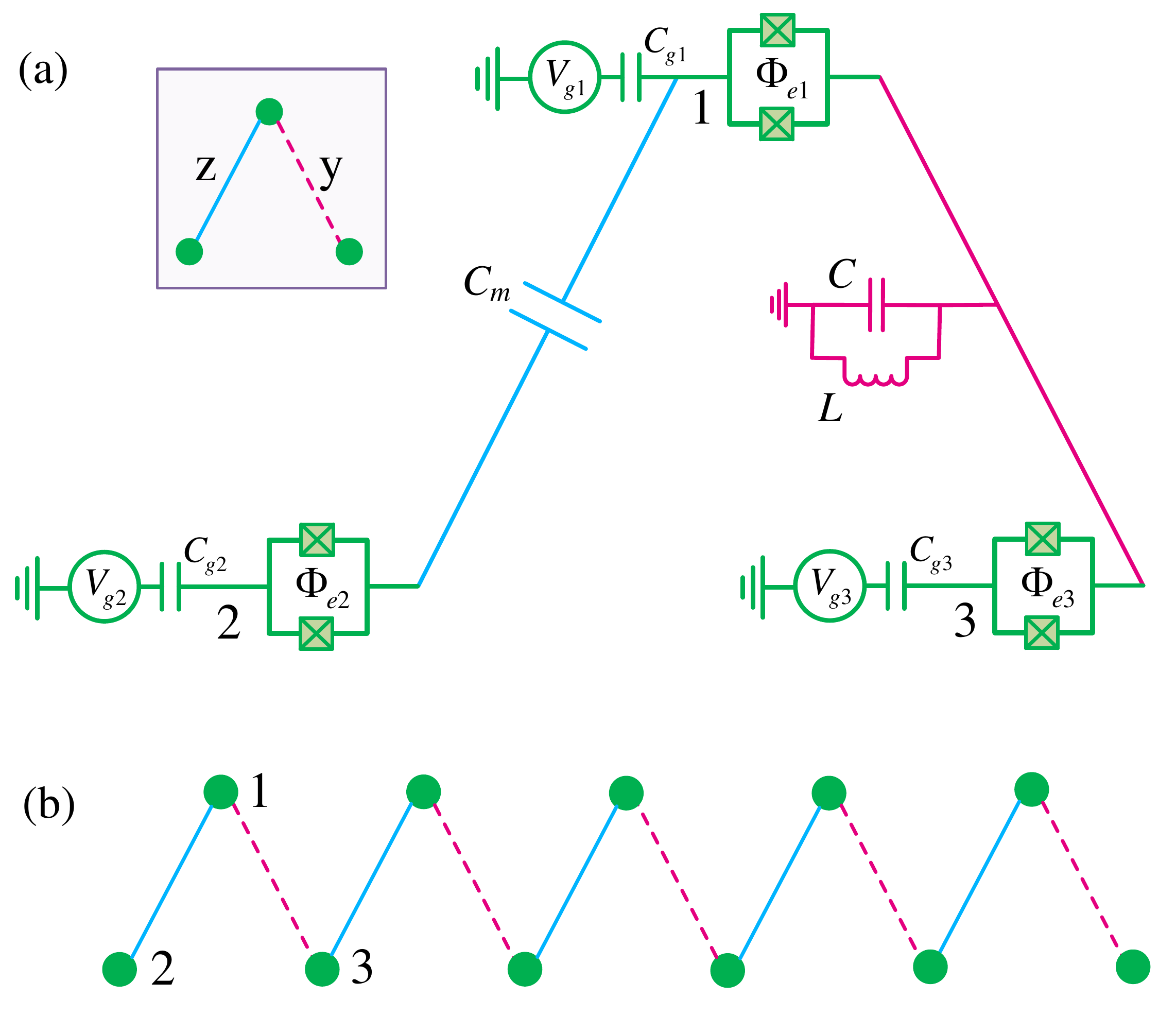}
\label{Fig:setup}
\caption{
Scheme of a circuit QED system for the physical implementation of the
1D compass chain Hamiltonian Eq. (\ref{com}):
(a)~Design of the basic building block, which is composed of three
superconducting charge qubits, labeled as $1,~2,~3$. Qubits 1 and 2
are coupled capacitively to each other via a mutual capacitance $C_m$;
and the coupling between the qubits 1 and 3 are provided by a
commonly-shared LC oscillator. Inset: the orange circles denote the
superconducting charge qubits; the two types of inter-qubit couplings
are denoted as $z$- and $y$-bond, which are indicated by the
blue-solid and the red-dashed line, respectively.
(b)~A~1D compass chain constructed by repeating the building block in
(a).}\label{Fig:setup}
\end{figure}

It has been demonstrated that charge qubits can be coupled to each
other for all the individual interactions of Ising type \cite{Wen05},
i.e., $\propto\sigma_i^x\sigma_{i+1}^x$ via a mutual inductance
\cite{you2002prl}, $\propto\sigma_i^y\sigma_{i+1}^y$ via an LC
oscillator \cite{Mak99}, and $\propto\sigma_i^z\sigma_{i+1}^z$ via a
capacitor \cite{pashkin2003n}. This provides us a promising way to
implement the 1D compass chain with the superconducting charge qubits.
As shown in Fig.~\ref{Fig:setup}, a charge qubit is placed at each
node, and is then connected to its two nearest neighbors with two
types of couplers, i.e., a capacitor for the $z$-type bond and an LC
oscillator for the $y$-type bond.

For the sake of simplicity and without the loss of generality, we
assume all the charge qubits to be identical such that
$C_{gi}\equiv C_g$, $E_{Ji}\equiv E_J$, $E_{Ci}\equiv E_C$. Following
the standard quantization procedure of the circuit by firstly writing
down the kinetic energy and the potential energy of the circuit, and
secondly choosing the average phase drop $\varphi_i$ of each charge
qubit as the canonical coordinate, the Hamiltonian of the entire system
can then be obtained by a Legendre transformation.

We consider the situation when the
frequency of the LC oscillator is much larger than the frequency of
the qubit. In this case the LC oscillator is not really excited and
the corresponding terms can be removed from the total Hamiltonian.
Even though, the LC oscillator's virtual excitation still produces an
effective coupling between the corresponding charge qubits. For charge
qubit with $E_C\gg E_J$, at very low temperature, the two-level
system is formed by the charge states $\ket{0}$ and $\ket{1}$, which
denote the zero and one extra Cooper pair on the island, respectively.
After projecting the total Hamiltonian into the $i$th charge qubit's
computational basis $\{\ket{0}_i,\ket{1}_i\}$, we obtain \cite{You10},
\begin{equation}
H=J_{i,j}^y\!\sum_{y-{\rm links}}\sigma_i^y\sigma_j^y
 +J_{i,j}^z\!\sum_{z-{\rm links}}\sigma_i^z\sigma_j^z
+\sum_i h_i^x \sigma_i^x,
\label{eq_htotal}
\end{equation}
where all the charge qubits are biased at the optimal point
(i.e., $n_{gi}=C_g V_{gi}/{(2e)=1/2}$) such that $h_i^z = 0$,
and $h_i^x= - E_J \cos(\pi \Phi_{ei}/\Phi_0)$ is the effective
Josephson energy of the $i$th charge qubit, $\Phi_0\equiv h/2e$
is the flux quantum. The $y$-type Ising coupling strength
\begin{eqnarray}
J_{i,j}^y=-4 \xi E_J^2 \cos\left(\pi \Phi_{ei}/\Phi_0\right)
\cos\left(\pi \Phi_{ej}/\Phi_0\right)\le 0,
\end{eqnarray}
with
$\xi=L\pi^2 {(2 C_J+C_g+C_m)}^2 {(C_g+C_m)}^2/{(\Lambda\Phi_0)}^{2}$,
are tunable via the external magnetic flux threading the SQUIDs in the
$i$th and $j$th charge qubits. Simultaneously the $z$-type coupling
strength is fixed as
\begin{eqnarray}
J_{i,j}^z=\frac{e^2 C_m}{\Lambda} \ge 0,
\end{eqnarray}
with $\Lambda=(2 C_J+C_g+C_m)^2-C_m^2$. A detailed analysis of circuit
quantization can be found in Ref. \cite{You10}.

An intuitive understanding of the coupling mechanism in the Hamiltonian
Eq.~(\ref{eq_htotal}) would be the following. Each charge qubit is
coupled to its left or right nearest neighbor via a capacitor or an LC
oscillator. The appearance of a capacitor modifies the electrostatic
energy of the system, and thus provides the $z$-type Ising coupling.
On the other hand, the magnetic energy of the inductor is biased by
a current composed of contributions from both of the two qubits,
and thus the virtually excited LC oscillator induces the $y$-type
Ising coupling. Then implementing a unitary rotation around the
$y$ axis, i.e., $U\equiv\prod_j e^{i\pi\sigma_j^z/4}$, one can find
$U\sigma_i^xU^\dagger=\sigma_i^z$, $U\sigma_i^zU^\dagger=-\sigma_i^x$,
and then Eq. (\ref{eq_htotal}) can be recast into the Hamiltonian Eq.
(\ref{com}).

\section{Effect of three-site interactions}
\label{sec:3site}

To make the model as general as possible and still exactly soluble,
we introduce in addition three-site interactions of the
(XZX$+$YZY)-type into Eq. (\ref{com}) ,
\begin{eqnarray}
\cal{H}_{\rm 3-site}&=&J_3\sum_{i=1}^{N}
\left(X_{i-1,i+1}+ Y_{i-1,i+1}\right)Z_{i},
\label{Hamiltonian5}
\end{eqnarray}
where $J_3$ characterizes their strength. Such interactions between
three adjacent sites emerge as an energy current of a compass chain in
the nonequilibrium steady states \cite{Qiu16}. Three-site interactions
violate the intermediate symmetry and elicit exotic phenomena. This
generalized version of the 1D QCM has been shown to host a diversity of
nontrivial topological phases and an emergent
BKT QPT under the interplay of a
perpendicular Zeeman field and multi-site interactions~\cite{You17}.

We next turn to the discussion of the physical implementation of the 1D
QCM including the three-site interaction with superconducting circuits.
As superconducting circuits offer advantages of easy tunability and
scalability, in principle, many-body interactions in superconducting
systems could be designed using Josephson-junction-based couplers in
a graph structure~\cite{bergeal2010np,chancellor2017nqi,leib2016qst}.
However, the effective many-body coupling terms may emerge with a much
weaker strength. An alternative and practical strategy to generate
many-body interactions would be the simulation protocols employing
the microwave fields with appropriate frequency conditions, as have
been studied in nuclear magnetic resonance systems~\cite{zhang2006pra},
optical lattices~\cite{buchler2007np}, and superconducting circuits~\cite{sameti2017pra,marvian2017pra}. Therefore, we expect
that the three-site interactions of the sort of XZX$+$YZY-type as in
Eq. (\ref{Hamiltonian5}) would be built in similar fashions. However,
an in-depth study of experimental implementation of this particular
model will be left for future investigation.

\begin{figure}
\includegraphics[width=\columnwidth]{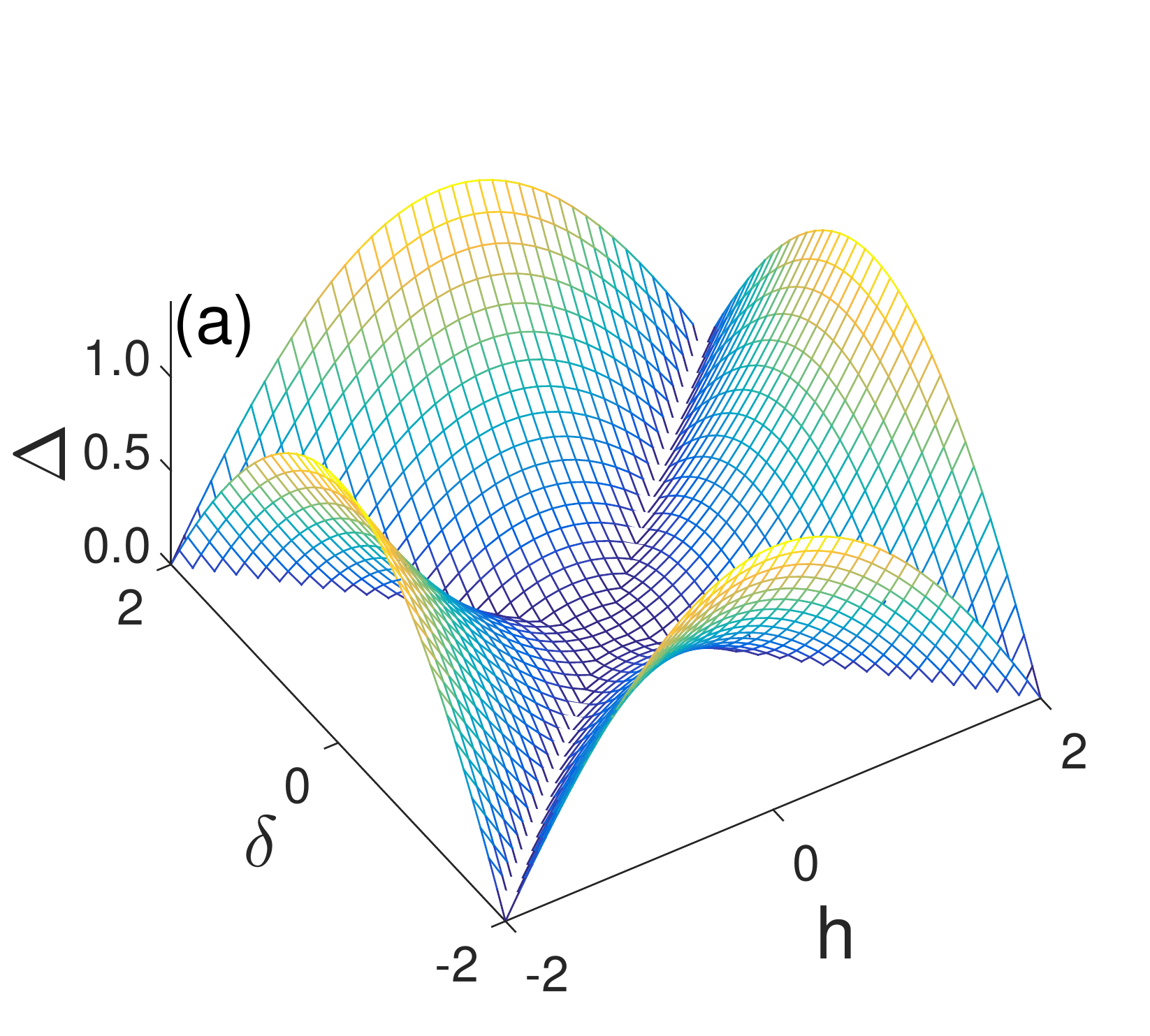}
\includegraphics[width=\columnwidth]{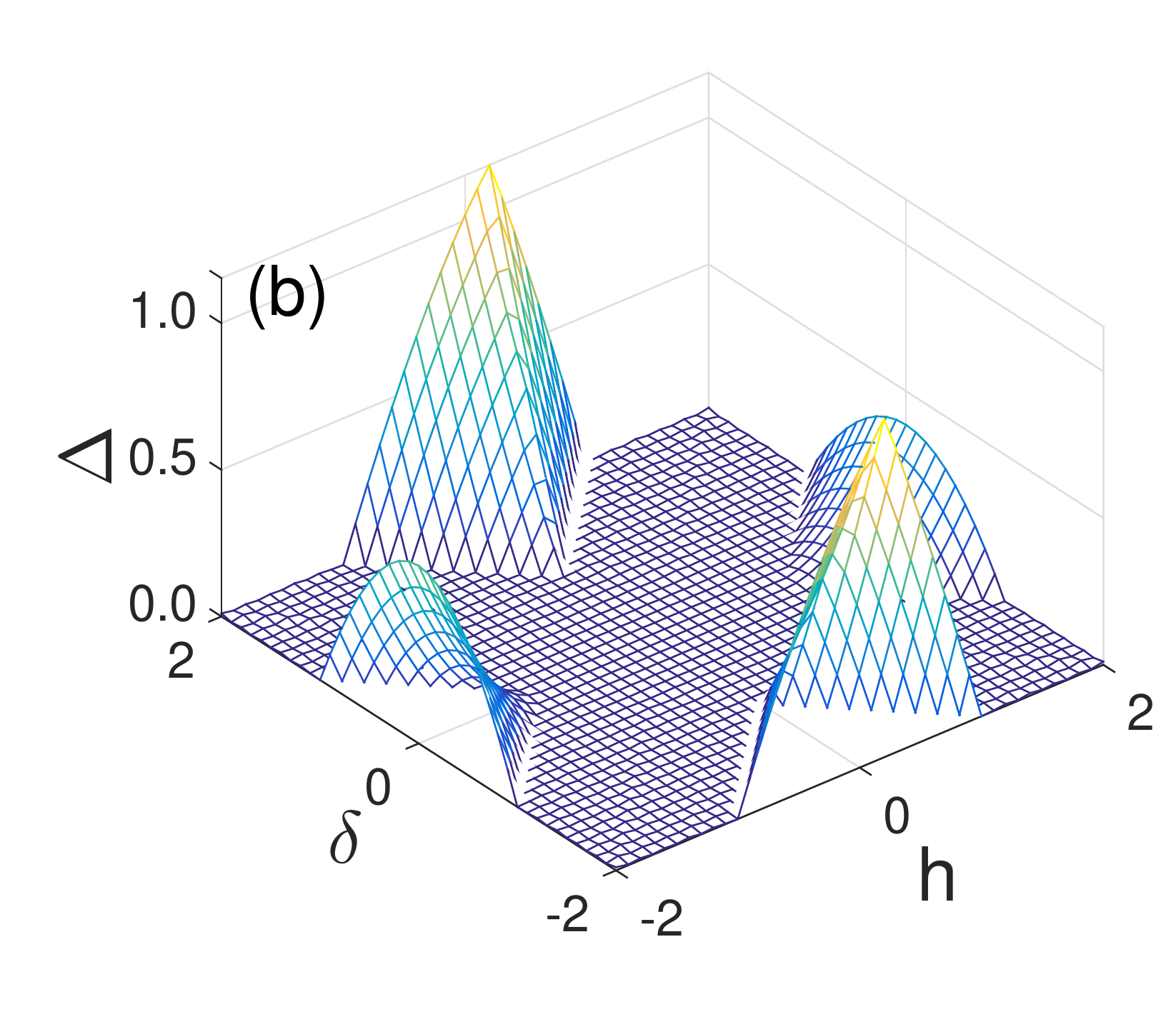}
\caption{Three-dimensional plot of the energy gap as a function of $h$
and $\delta$ for:
(a) $J_3=0$ and
(b) $J_3=1$.
Parameters are as follows: $J_1=1$, $J_2=4$.}
\label{Fig:Gap}
\end{figure}

\begin{figure*}
\includegraphics[width=1.0\textwidth]{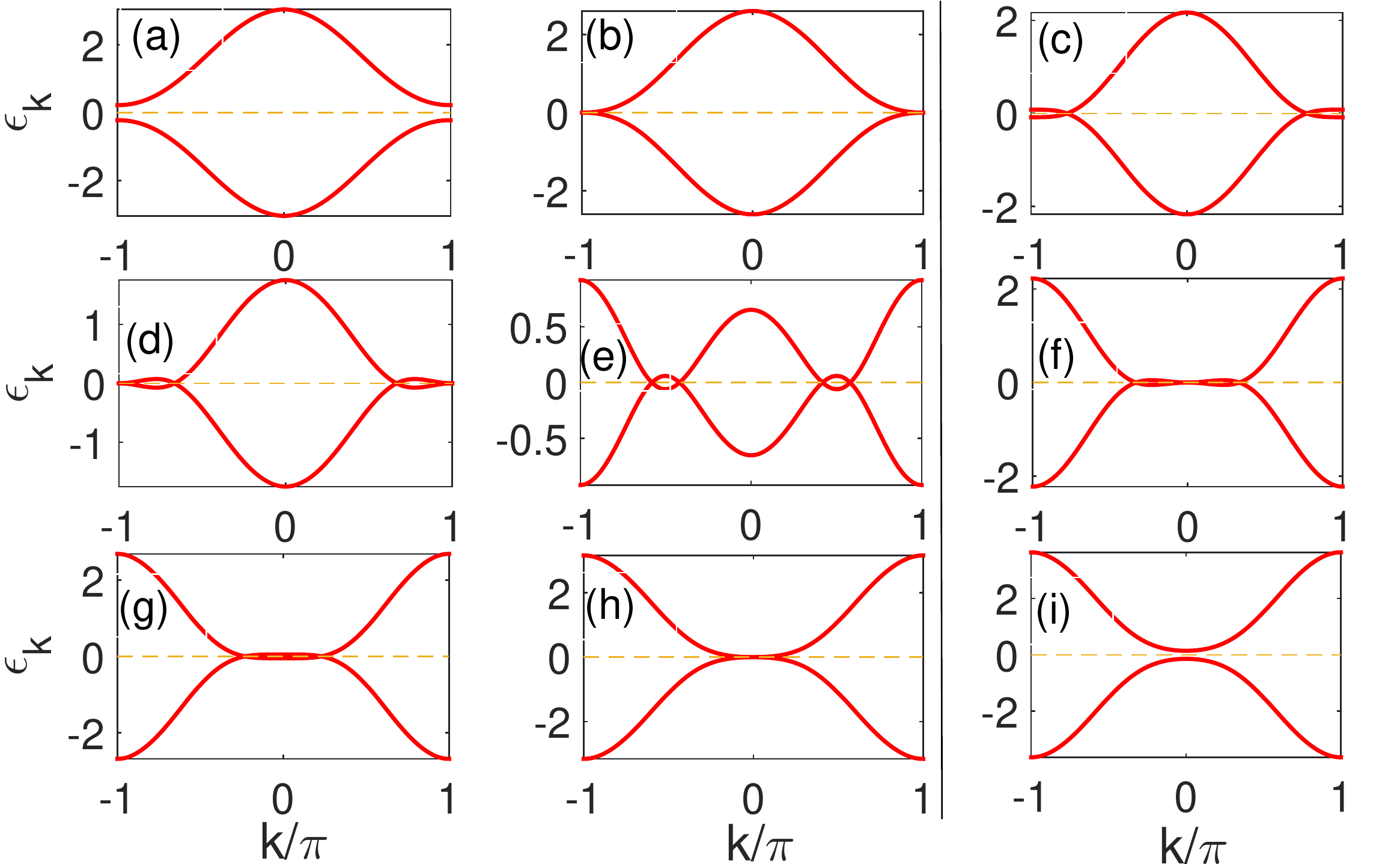}
\caption{The electron spectra $\varepsilon_{k,2}$ and its corresponding
hole spectra $-\varepsilon_{k,2}$ for increasing magnetic field $h$ in
Eq. (\ref{model}):
(a)~$h=-3$, (b) $h=-2.5$, (c) $h=-2.0$, (d) $h=-1.5$, (e) $h=0$,
(f) $h=1.5$, (g) $h=2$, (h) $h=2.5$, and (i) $h=3$.
Parameters are as follows:  $J_1=1$, $J_2=4$, $J_3=2$, and $\delta=0.5$. }
\label{Fig:Spectrum_delta}
\end{figure*}

The generalized Hamiltonian of the 1D QCM which includes the three-site
(XZX$+$YZY) interaction is
\begin{eqnarray}
{\cal H}=\cal{H}_{\rm QCM}+\cal{H}_{\rm h}+\cal{H}_{\rm 3-site}.
\label{model}
\end{eqnarray}
Eq. (\ref{model}) describes a 1D $sp$-chain with inter-band
interactions and hybridization between orbitals \cite{Puel15}.
The three-site interactions can be converted into fermionic form
${\cal H}_{\rm 3-site}= 4 J_3\sum_k \cos k\,c_k^\dagger c_k$.
We note that the spectra can be pinpointed at commensurate momenta
$k=\pm\pi/2$ regardless of the value of $J_3$. Hence the eigenspectra
(\ref{excitationspectrum2}) can be renormalized with
$-2h\to F_k = -2h+2J_3\cos k$, as evidenced in Eq. (\ref{APP:spec}).

The main features and the evolution of these profiles under staggered
fields with increasing magnetic field $h$ are depicted in Fig.
\ref{Fig:Spectrum_delta}. We observe that the ground state of the
system is complicated under the interplay of three-site interactions
and staggered magnetic field. As $h$ rises from large negative values,
$\varepsilon_{k,2}$ closes the gap gradually and finally touches
$\varepsilon=0$ at momentum $k=\pi$ for $h=-\vert J_3+\delta \vert$.
Further increase of $h$ bends $\varepsilon_{k,2}$ downwards, leading
to $\varepsilon_{\vert k\vert >\vert k_{ic}\vert,2}<0 $ with an
incommensurate momentum $k_{ic}$. An additional crossing at $k=\pi$
revives for $h=-\vert J_3-\delta\vert$. We can see that the number of
crossing points at zero energy grows from 0 to 4 in Figs.
\ref{Fig:Spectrum_delta}(a)-\ref{Fig:Spectrum_delta}(e), and then
decreases with further increase of $h$, see Figs.
\ref{Fig:Spectrum_delta}(f)-\ref{Fig:Spectrum_delta}(i). Altogether,
the number of Fermi points at which the linear dispersion relation is
found changes as $h$ increases. Indeed, here the topological transition
belongs to the universality of the Lifshitz transition.

\begin{figure}[t!]
\includegraphics[width=\columnwidth]{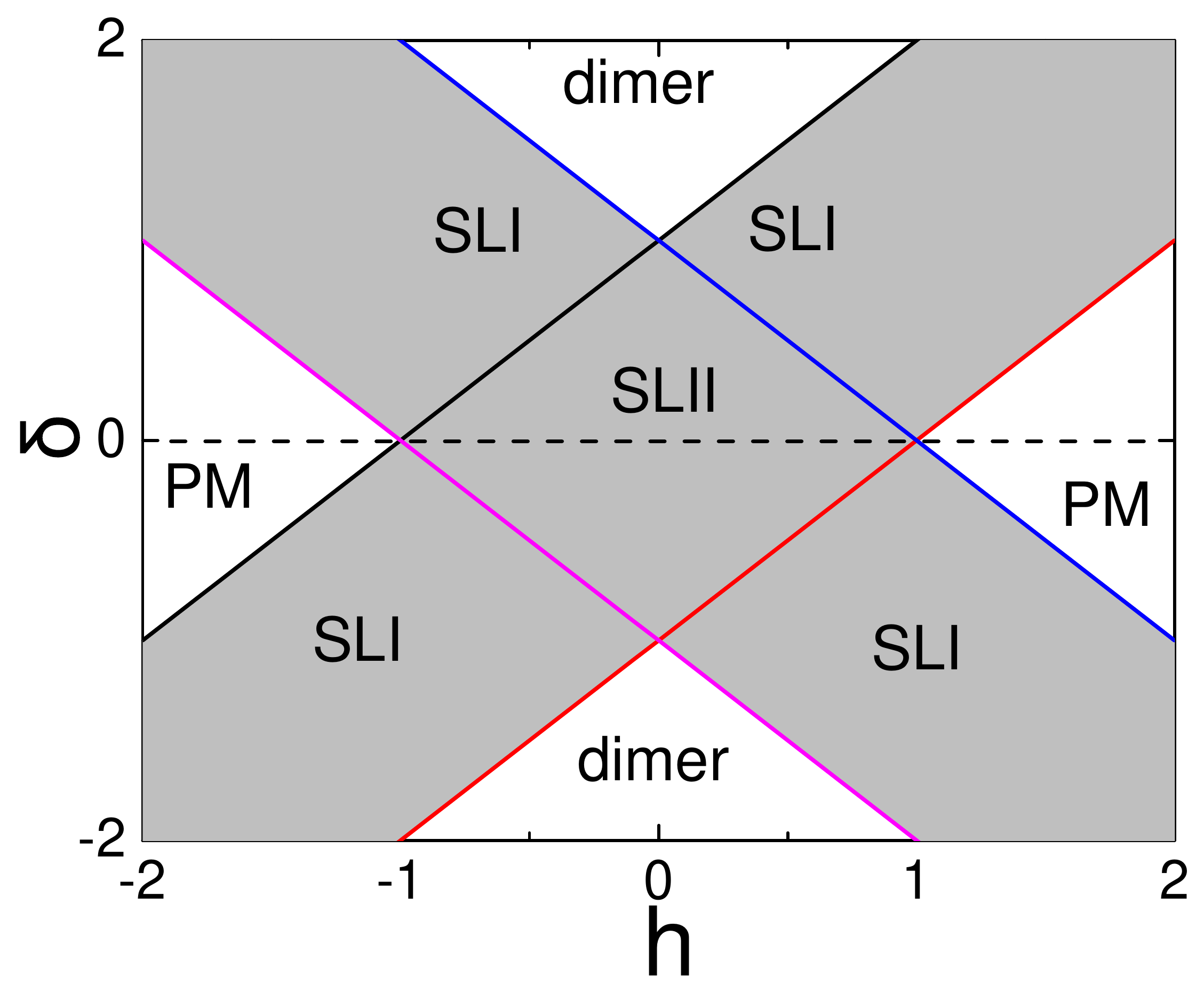}
\caption{Phase diagram of the compass chain (\ref{model}) in an
alternating transverse field. The shaded regions mark the gapless
spin-liquid phase, in which SLI (SLII) denotes the spin-liquid phase I
(II) with 2 (4) Fermi nodes. The dashed line marks the path $\delta=0$.
Other parameters: $J_1=1$, $J_2=4$, $J_3=1$.  }
\label{Fig:sigmaz_PD}
\end{figure}

It is also easy to see that the Weyl points collapse at
$h=\pm\vert J_3+\delta\vert$ with increasing $h$. In the gapless phase
the crossings between bands exhibit a linear dispersion relation [see
Figs. \ref{Fig:Spectrum_delta}(c)-\ref{Fig:Spectrum_delta}(g)] and
thus define effective 1D Weyl modes. One notes that the nodes appear
and disappear only when two nodes are combined, as a characteristic of
Weyl fermions in a three-dimensional or 2D superconductor
\cite{Yang14,Neu16} and in topological superfluidity
\cite{Cao14,Pav15,Brz17,Brz18}. It is noticed with the emergence of two
Weyl points at its extremities ($k=\pm\pi$) and their collapse at the
center of the Brillouin zone ($k=0$). In this region the low-energy
Hamiltonian with Weyl nodes in a 1D system can be reduced
to describe the two Bogoliubov bands that cross zero energy.

The resulting phase diagram of the model Eq. (\ref{model}), obtained by
the exact solution using the Jordan-Wigner transformation, is shown in
Fig. \ref{Fig:sigmaz_PD}. For clarity we have considered the entire
plane of fields, although the phase diagram is obviously symmetric
under reflection from the $h_1=h_2$ and $h_1=-h_2$ lines, which also
symmetrize the spectrum and entanglement. The quantum phase boundaries
are determined by the following condition:
\begin{equation}
\vert h \vert +\vert \delta \vert= J_3.
\end{equation}
For large $h$, the system changes to a disordered paramagnetic (PM)
phase, in which the $z$-axis sublattice magnetizations are unbiased,
as shown in Fig. \ref{magnetization}(a).
On the contrary, the dimer phase is the one in which
$z$-axis magnetization at odd and even sites has a staggered order.

The staggered magnetic susceptibilities are vanishing in the both
gapped phases, while they are finite in the gapless phases. Besides,
as shown in Fig. \ref{magnetization}(b), the
nearest neighbor
correlation $\langle\sigma_{2i-1}^y\sigma_{2i}^y\rangle$ clearly shows
nonanalytical behavior at the QCPs, and the counterpart
$\langle\sigma_{2i-1}^x\sigma_{2i}^x\rangle$ is smooth. One can notice
that a singular behavior can be detected by taking the first
derivative of $\langle\sigma_{2i-1}^x \sigma_{2i}^x\rangle$ and
$\langle\sigma_{2i}^y\sigma_{2i+1}^y\rangle$ with respect to $h$.

Since QPTs are caused by nonanalytical behavior of ground-state energy,
QCPs correspond to zeros of $\varepsilon_{k,2}$. The gap vanishes as
$\Delta\sim(h-h_c)^{\nu z}$, where $\nu$ and $z$ are the
correlation-length and dynamic exponents, respectively.
The gap is determined by the condition,
$\partial\varepsilon_{k=k_0,2}/\partial k=0$ and one finds
$\Delta =\min_{k}\vert\varepsilon_{k,2}\vert$.
This implies that the minimum is suited at either $k_0=0$ or $k_0=\pi$,
depending which mode has a lower energy. One finds the critical
exponents satisfy $\nu z=1$, as revealed in Fig. \ref{Fig:Gap}(b).
The expansion of the energy spectra at the criticality around the
critical mode $k_0$, i.e., at $\Delta k \equiv k-k_0 \ll 1$, \break
\noindent $\partial\varepsilon_{k,2}\sim
2J_3\delta(\Delta k)^2/\sqrt{(J_1+J_2\cos k_0)^2+4\delta^2}$.
The quadratic dispersion in Fig. \ref{Fig:Spectrum_delta}(b) suggests
a dynamical exponent $z=2$ and hence $\nu=1/2$, which is different
from the generic QCM in the absence of three-site interactions
\cite{Sun09a,Motamedifar13}.

Remarkably, the ground state develops weak singularities at $\delta=0$.
For $\delta=0$ the phase boundaries are pinpointed at $h_c=\pm J_3$
with an incommensurate critical momentum $k_0=\cos^{-1}(h/J_3)$, as
presented in Fig. \ref{Fig:Gap}(b). One can find that the system
transforms from the gapped phase to the gapless phase passing through
an unconventional field-induced QCP, where infinite-order QPTs occur by
tuning $h$ along the path (Fig. \ref{Fig:sigmaz_PD}, dashed line) to
approach the QCPs, with no broken-symmetry order parameter.
We can identify that $h_c=\pm J_3$ are multi-critical points, where
$h-\delta=\pm J_3$ and $h+\delta=\pm J_3$ meet \cite{You17}. One finds
the critical exponents that follow $\nu z=2$ by observing the gap
scaling. The dependence of low-energy excitations on $k$ shows that
$z=2$ in the gapless phase while $z=4$ at QCPs. It has been shown that
$z$ can  be extracted from the measurement of the low-temperature
specific heat and entropy in the Tomonaga-Luttinger liquid phase
\cite{Kono15}.

\section{Quantum coherence measures}
\label{sec:measures}

In the representation spanned by the two-qubit product states we employ
the following basis,
\begin{equation}
\{\vert 0\rangle_i\otimes\vert 0\rangle_j,
  \vert 0\rangle_i\otimes\vert 1\rangle_j,
  \vert 1\rangle_i\otimes\vert 0\rangle_j,
  \vert 1\rangle_i\otimes\vert 1\rangle_j\},
\end{equation}
where $\vert 0\rangle$ ($\vert 1\rangle$) denotes spin up (down) state,
and the two-site density matrix can be expressed as,
\begin{equation}
\rho_{ij}=\frac{1}{4}\sum_{a,a'=0}^3\left\langle
\sigma_i^{a}\sigma_j^{a'}\right\rangle\,\sigma_i^{a}\sigma_j^{a'},
\end{equation}
where $\sigma_i^a$ stands for Pauli matrices
$\{\sigma_i^x,\sigma_i^y,\sigma_i^z\}$ with $a=1,2,3$, and for a
$2\times 2$ unit matrix with $a=0$. Since the Hamiltonian has
$\mathbb{Z}_2$ global phase-flip symmetry, the two-qubit density
matrix reduces to an $X$ state,
\begin{equation}
\rho_{ij}=\left(
\begin{array}{cccc}
u_{+} & 0 & 0 & z_{-} \\
0 & w_{+} & z_{+} & 0 \\
0 & z_{+}^{*} & w_{-} & 0 \\
z_{-}^{*} & 0 & 0 & u_{-}
\end{array}\right),
\label{eq:2DXXZ_RDM}
\end{equation}%
with
\begin{eqnarray}
u_{\pm }&=&\frac{1}{4}\left(1 \pm \langle {\sigma_{i}^{z}}
\rangle \pm \langle {\sigma_{j}^{z}}\rangle
+\langle {\sigma _{i}^{z}\sigma _{j}^{z}}\rangle\right),
\label{upm} \\
z_{\pm}&=&\frac{1}{4}\left(\langle \sigma _{i}^{x}\sigma _{j
}^{x}\rangle \pm \langle \sigma_i^y\sigma_j^y\rangle\right),
\label{z1} \\
\omega _{\pm}&=& \frac{1}{4}\left(1 \pm
\langle {\sigma_{i}^{z}}\rangle \mp
\langle {\sigma_j^z}\rangle-\langle\sigma_i^z\sigma_j^z\rangle\right).
\label{omega1}
\end{eqnarray}

\begin{figure}[t!]
\includegraphics[width=8cm]{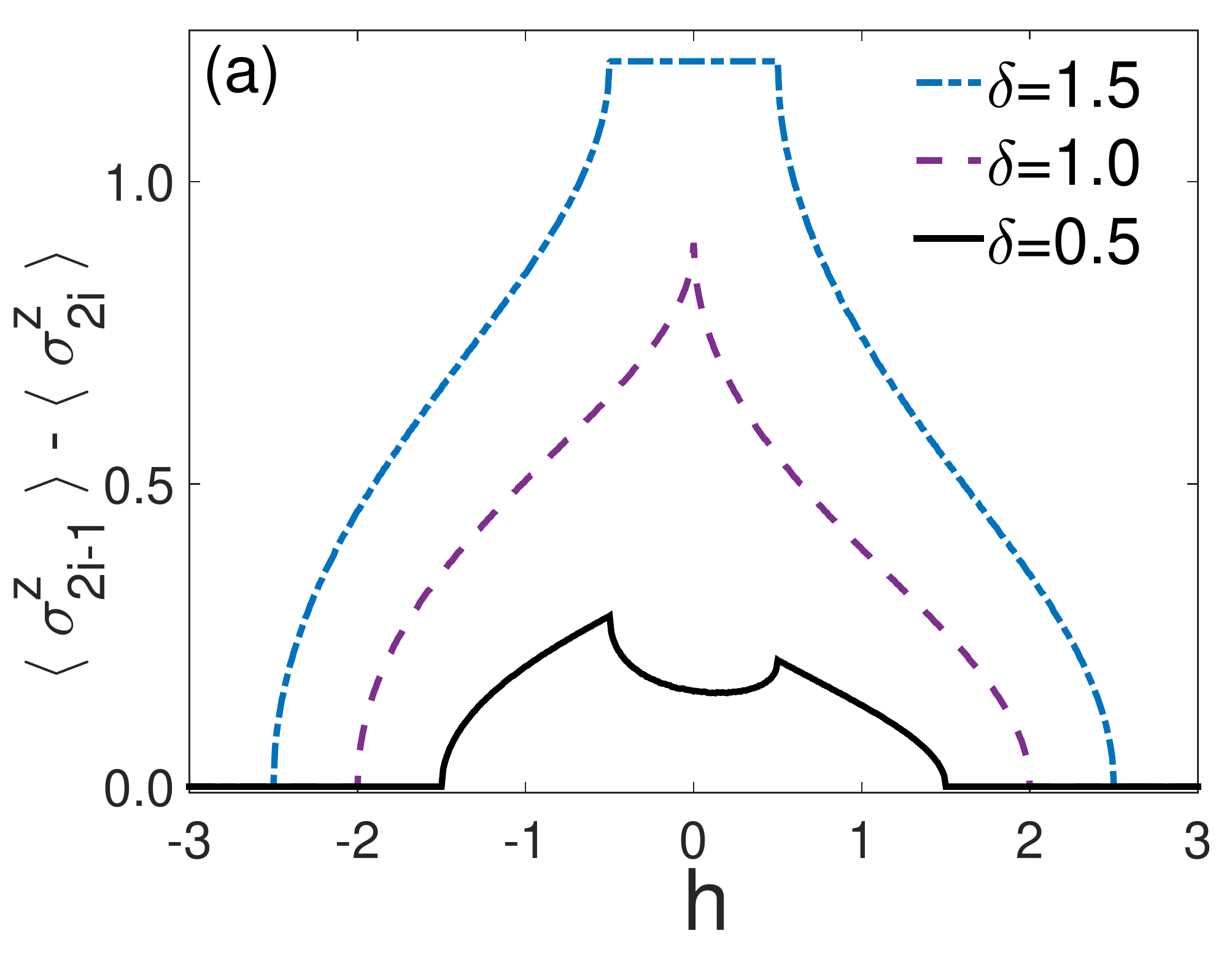}
\includegraphics[width=8cm]{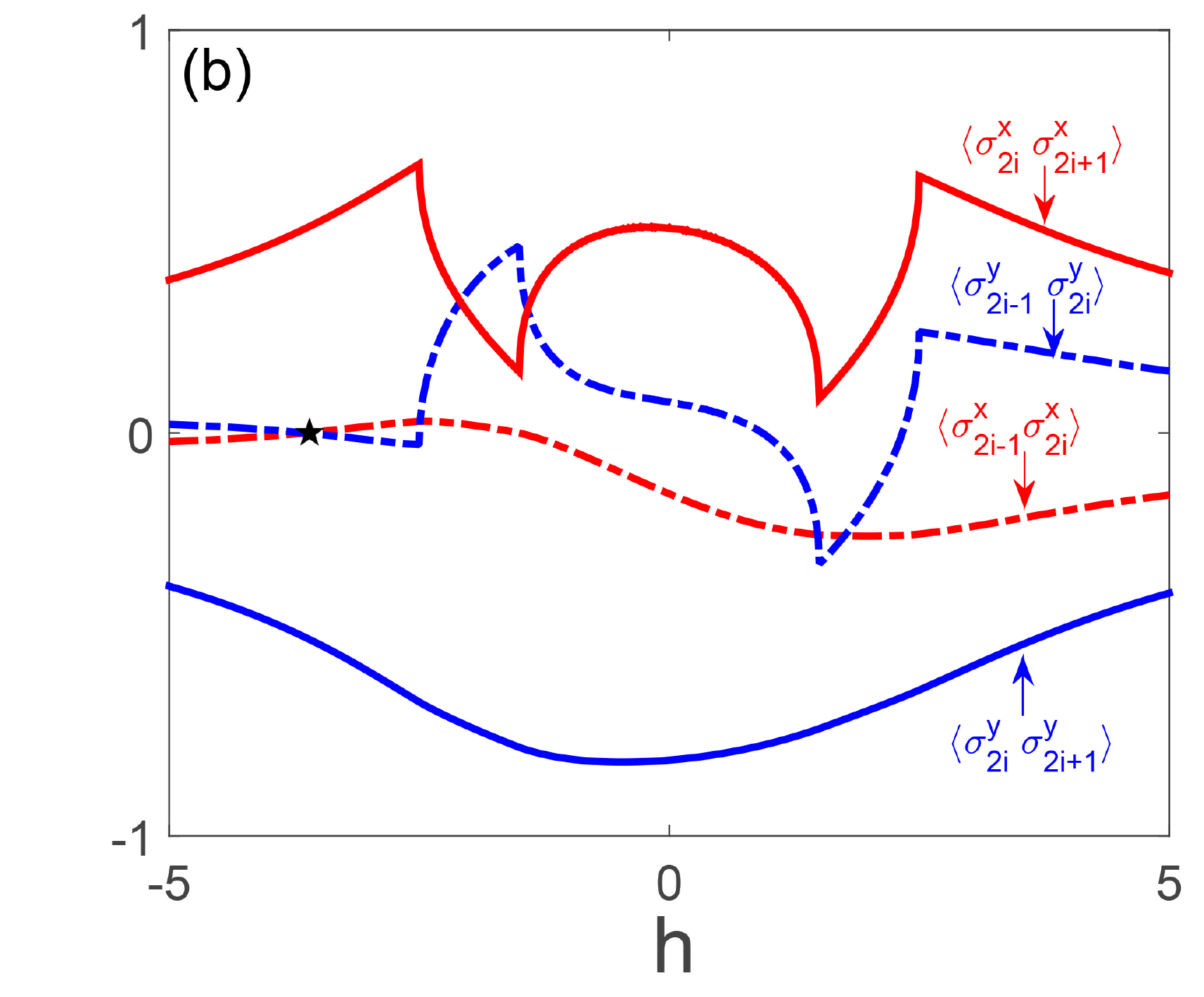}
\caption{Effect of alternating magnetic field in the 1D compass
model (\ref{model}):
(a) the difference between odd-site and even-site magnetization
for $\delta=0.5$, 1.0 and 1.5 with $J_3=1$;
(b) the nearest-neighbor correlations
$\langle\sigma_{i}^\alpha \sigma_{i+1}^\alpha\rangle$ ($\alpha=x$, $y$)
on odd bonds and even bonds with $J_3=2$, $\delta=0.5$.
The symbol $\star$ in (b) marks the position of
$\langle\sigma_{2i-1}^x \sigma_{2i}^x\rangle
=\langle\sigma_{2i-1}^y \sigma_{2i}^y\rangle=0$.
Other parameters are: $J_1=1$, $J_2=4$.}
\label{magnetization}
\end{figure}

Note that the formula can be simplified when the system is translation
invariant, i.e.,
$\langle\sigma_{i}^z\rangle=\langle\sigma_{j}^z\rangle$ for arbitrary
$i$ and $j$, such that $\omega_{+}=\omega_{-}$. Under the staggered
magnetic field, the magnetization densities
$\{\langle\sigma_{2i-1}^z\rangle\}$ at odd sites and
$\{\langle\sigma_{2i}^z\rangle\}$ at even sites are inequivalent.
As is disclosed in Fig. \ref{magnetization}(a), the difference of the
$z$-axis magnetizations is nonvanishing in the gapless regions and
dimer phase. By means of the Wick theorem, it is well known that
two-site correlation functions can be expressed as an expansion of
Pfaffians \cite{EBarouch70,Osb02}. One easily finds that
\begin{eqnarray}
S(\rho_{ij})=-\sum_{m=0,1}\xi_m\log_2\xi_m-\sum_{n=0,1}\xi_n\log_2\xi_n,
\end{eqnarray}
where
\begin{eqnarray}
\xi_m &=&  \frac{1}{4}\left\{1+ \langle
\sigma_{i}^z \sigma_{j}^z \rangle  +  (-1)^m
\left[\left(\langle\sigma_i^x \sigma_j^x\rangle-\langle\sigma_i^y
\sigma_j^y\rangle\right)^2 \right.\right. \nonumber \\
& +&\left.\left.\left(\langle
\sigma_i^z\rangle+\langle\sigma_j^z\rangle\right)^2 \right]^{1/2}\right\},
\\
\xi_n &=& \frac{1}{4}\left\{1 - \langle \sigma_{i}^z \sigma_{j}^z \rangle
+(-1)^n \left[\left(\langle\sigma_i^x \sigma_j^x\rangle+\langle\sigma_i^y
\sigma_{j}^y\rangle\right)^2 \right.\right. \nonumber \\
&+&\left.\left.\left(\langle\sigma_i^z\rangle-\langle\sigma_j^z \rangle\right)^2\right]^{1/2}\right\}.
\end{eqnarray}

Recently diagonal discord was proposed to be an economical and
practical measure of discord \cite{Liu17}, which compares quantum
mutual information with the mutual information revealed by a
measurement that corresponds to the eigenstates of the local density
matrices. As long as the local density operator is nondegenerate,
diagonal discord is easily computable without optimization over all
possible local measurements, which make the discord-like quantities
unamiable. The reduced density matrix for a single $i$th-qubit has a
local eigenbasis $\prod_\eta=\vert\eta\rangle\langle\eta\vert$ when
$\langle\sigma_{i}^z\rangle\neq 0$, and then a local measurement
follows $\pi_i(\rho_{ij})=
\sum_\eta (\prod_\eta\otimes I_j)\rho_{ij}(\prod_\eta\otimes I_j)$.
Diagonal discord $\bar{D}_i(\rho_{ij})$ characterizes the reduction
in mutual information induced by $\pi_i(\rho_{ij})$ and takes a
similar form to the relative entropy as
$\bar{D}_i(\rho_{ij})=S(\rho_{ij}\vert\vert\pi_i(\rho_{ij}))$.

\begin{figure}[t!]
 \includegraphics[width=\columnwidth]{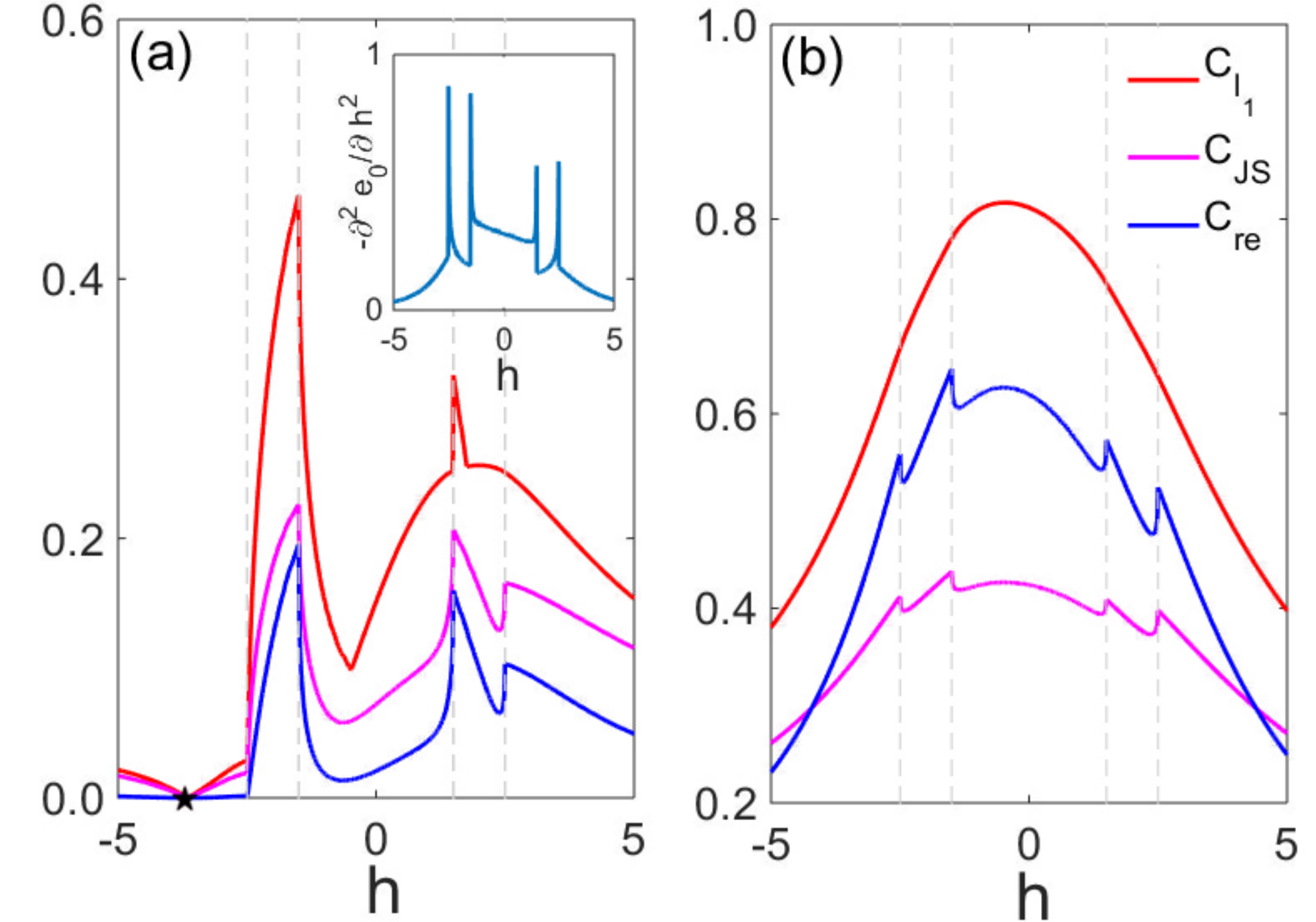}
\caption{
Quantum coherence measures on (a) odd bonds and (b) even bonds
for increasing magnetic field $h$ with $J_1=1$, $J_2=4$, $J_3=2$,
$\delta=0.5$. The legend shown in (b) is the same with (a).
The symbol $\star$ marks the position of $h_\star=-3.647$.
The dashed lines correspond to positions of $\pm\vert J_3\pm\delta\vert$.
Inset in (a) shows the second-order derivative of the ground-state
energy $e_0$.}
\label{Correlation3}
\end{figure}

\begin{figure}[t!]
 \includegraphics[width=\columnwidth]{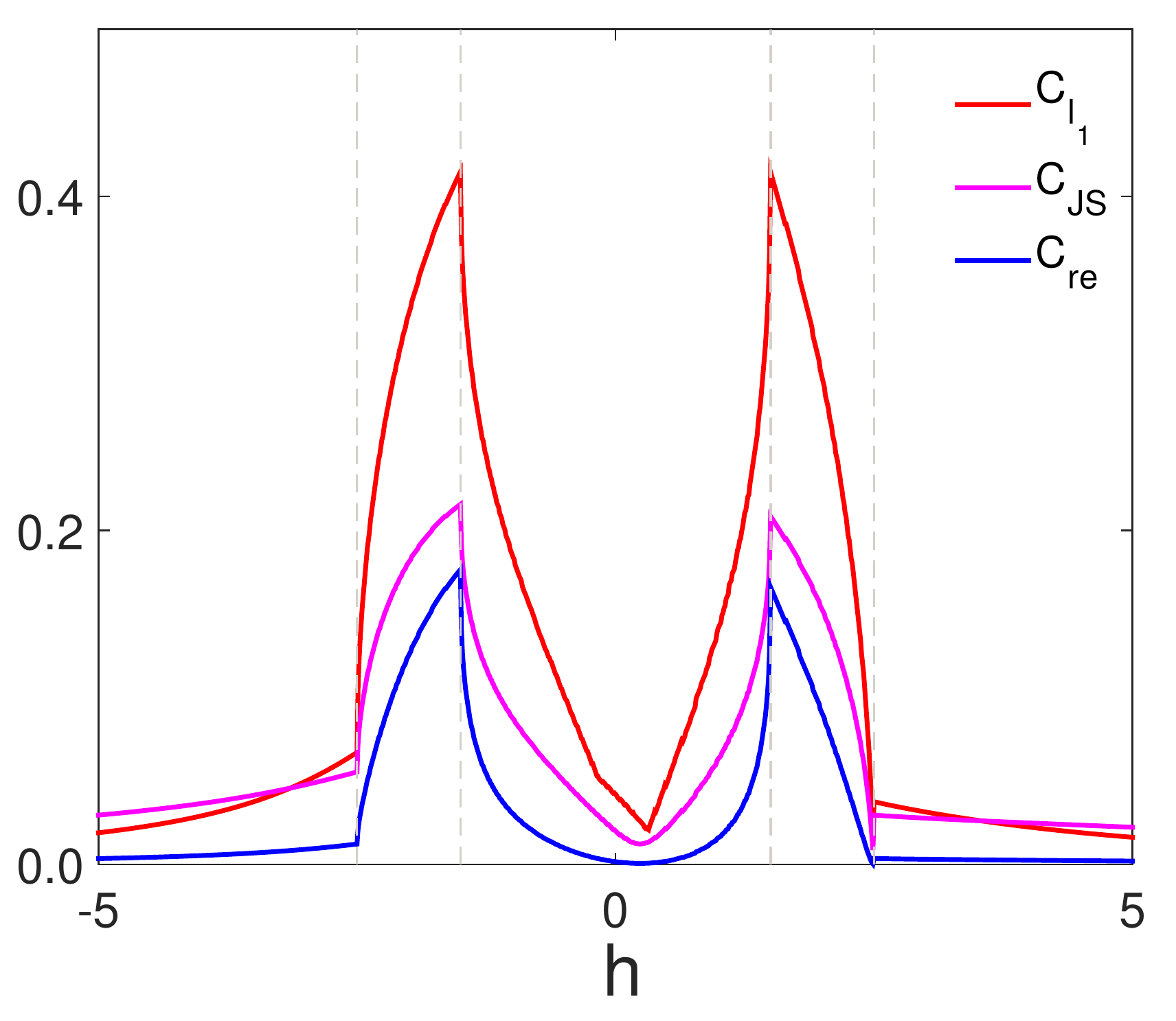}
\caption{
Quantum coherence measures of next nearest neighbor qubits for
increasing magnetic field $h$
with $J_1=1$, $J_2=4$, $J_3=2$, $\delta=0.5$.
The dashed lines mark the positions of $\pm\vert J_3\pm\delta\vert$.}
\label{Correlation4}
\end{figure}

In terms of the two-qubit $X$ state in Eq. (\ref{eq:2DXXZ_RDM}),
$\bar{D}_i( \rho_{ij} )$ is identical to $C_{\rm re}(\rho_{ij})$.
Without the loss of generality, we mainly use $C_{\rm re}(\rho)$
hereafter although it has two-fold implications in quantum correlations.
Besides, the $l_1$ norm quantum coherence can be simplified to:
\begin{eqnarray}
C_{l_1}(\rho)&=&\max\left\{
\left\vert\langle\sigma_i^x\sigma_j^x\rangle\right\vert,
\left\vert\langle\sigma_i^y\sigma_j^y\rangle\right\vert\right\}.
\label{l1norm_explicitform}
\end{eqnarray}
For our purposes, the correlation and coherence measures in the ground
state of the quantum compass chain under an alternating transverse
magnetic field for two spins are investigated in the following for
comparison. Figures \ref{Correlation3} and \ref{Correlation4} display
the results for the relative entropy, the $l_1$ norm quantum coherence
and the JS divergence of two qubits in the ground state of the compass
chain. We find $C_{l_1}(\rho)\ge C_{re}(\rho)$ holds, as was
conjectured in Ref. \cite{Rana16}. The conjecture was proved only for
the pure state and not yet proven for mixed state, and our findings
indicate its validity.

Looking at the relative entropy, the $l_1$ norm quantum coherence and
the JS divergence on the odd bonds presented in Fig.
\ref{Correlation3}(a), three measures become zero simultaneously at
$h_\star=-3.647$. The relative entropy shows a smooth local minimum
at this special point, which is different from nonanalytical behaviors
of its counterparts. The null point of coherence measures corresponds
to a factoring point, where the intersite-correlators on the weak bonds
$\langle\sigma^x_{2i-1}\sigma^x_{2i}\rangle$ and
$\langle\sigma^y_{2i-1}\sigma^y_{2i}\rangle$ vanish,
see Fig. \ref{magnetization}(b).
As a consequence, the density matrix (\ref{eq:2DXXZ_RDM}) becomes
diagonal in the orthogonal product bases. This factoring point is
found to be $\delta$ independent, and such a accidental inflexion will
be absent in the quantum coherence measures for the even bonds,
see Fig. \ref{Correlation3}(b), and the next-nearest-neighboring qubits,
see Fig. \ref{Correlation4}. One finds the quantum coherence on even
bonds is larger than that on odd bonds, and the next-nearest neighbor
coherence is a little smaller.

For $\delta=0$ the system undergoes a BKT phase transition, and the
quantum coherence exhibits a local maximum; see Fig. 10(a) in Ref.
\cite{You17}. For $\delta\neq 0$, the transitions for increasing $h$
belong to second-order phase transitions, as is verified in inset of
Fig. \ref{Correlation3}(a). In this respect, we find that the coherence
measures exhibit either a nonanalytical behavior or an extremum across
QCPs, indicating sudden changes take place.

After a closer inspection we find the $l_1$ norm quantum coherence
displays anomalies at QCPs $h_c=-J_3-\delta$, $\pm(J_3-\delta)$, but
it misses the QCP at $h_c=J_3+\delta$. We also observe that there are
superfluous kinks of the $l_1$ norm of coherence in regions around
$h=0$, as shown in Figs. \ref{Correlation3}(a) and \ref{Correlation4}.
The artificial turning points can be ascribed to the definition of
$l_1$ norm quantum coherence in Eq. (\ref{eq:measure1}). Further, this
norm does not exhibit any anomaly on the strong bonds, as shown in
Fig. \ref{Correlation3}(b). From such a comparison one can find that
relative entropy [Eq. (\ref{eq:measure2})] can faithfully reproduce
quantum criticality. Despite of their formal resemblances,
the JS divergence and the $l_1$ norm quantum coherence,
as different perspectives of quantum coherence, embody infidelity of
density matrix and are insufficient to readout the locations of QCPs.

\section{Discussion and summary}
\label{sec:con}

In this paper we consider a one-dimensional Hamiltonian with
short-range interactions that includes three-site interactions and
alternating magnetic fields. The one-dimensional quantum compass model
is a paradigmatic scenario of quantum many-body physics, which is more
subtle than the Ising model, and hence hosts richer phase diagrams.
For a second-order quantum phase transition from a gapped N\'{e}el
phase to a gapped paramagnetic phase, tools of quantum information
theory can always be employed to characterize successfully the
transition points. Usually achieving a complete and rigorous
quantum-mechanical formulation of a many-body system as desired is
obstructed by the complexity of quantum correlations in many-body
states.

The spin chain in the present model is efficiently solvable using the
standard Jordan-Wigner and Bogoliubov transformation techniques.
Adopting the exact solvability we describe the phase diagram of the
model as a function of its parameters. The perpendicular Zeeman field
and and three-site interactions spoil the intermediate symmetry in the
generic compass model, and thus they destroy the ground-state degeneracy of
the quantum compass chain. The tunability of the staggered magnetic
field entails the ground state can be among paramagnetic phase, dimer
phase, and spin-liquid phases, in which the number of Fermi points
falls into two categories. Except the multi-critical points, the phase
transitions are of second order. The critical exponents can be
extracted from low-energy spectra and gap scalings. Our investigations
show a uniform magnetic field can drive the spin-liquid phase to the
paramagnetic phase through the Berezinskii-Kosterlitz-Thouless
transition, where a multi-critical point is found.

Since a quantum phase transition is driven by a purely quantum change
in the many-body ground-state correlations, the notion of quantum
coherence appears naturally and is suited to probe quantum criticality.
To this end, we provide a study of associated exhibited quantum
correlations in this model using a variety of quantum information
theoretical measures, including the relative entropy and the $l_1$ norm
quantum coherence along with the Jensen-Shannon divergence. These
alternative frameworks of coherence theory stem mainly from different
notions of incoherent (free) operations. It is thus desirable and
interesting to find any interrelation between them.

We then compare the respective kinds of insights that they provide.
We have found that the continuous phase transitions occurring in this
model can be mostly faithfully detected by examining quantum
information theoretical measures. We also discern some differences.
The $l_1$ norm quantum coherence (defined as the sum of absolute values
of off diagonals in the reduced density matrix) of odd bonds develops
a singular behavior at non-criticality, which is caused by the absolute
operator in the definition (\ref{l1norm_explicitform}).
Moreover, the $l_1$ norm quantum coherence of even bonds does not
exhibit any anomaly across the critical points. A closer inspection
reveals that the $l_1$ norm quantum coherence of even bond shows an
inflection point and the transition point can be easily captured
looking at its derivative with respect to $h$, which would display
an extremum. On the contrary, the relative entropy and the
Jensen-Shannon divergence show pronounced anomalies, either a sharp
local maximum or a turning point. That is to say that they faithfully
sense the rapid change of quantum correlation, resulting in a clear
identification of quantum phase transitions.
Also the $l_1$ norm quantum coherence and the Jensen-Shannon divergence
become nonanalytical at exception points. Despite formal similarity,
different measures of quantum coherence have their respective scope for
detecting the quantum criticality. From such comparison, we believe
that the relative entropy is more credible than others.

Summarizing, our results
suggest that the diagonal entries of the density operator are
indispensable to extract information across the quantum critical
points. In other words, quantum phase transitions are cooperative
phenomena where competing orders induce qualitative changes in
many-body systems. The figures of merit of these measures might
be crucial to the optimizing basis. Furthermore, we proposed an
experimental scheme using superconducting quantum circuits to
realize the compass chain with alternating magnetic fields.

\acknowledgments

We thank Wojciech Brzezicki for insightful discussions.
W.-L. Y. acknowledges NSFC under Grant Nos. 11474211 and 61674110.
Y. Wang acknowledges China Postdoctoral Science Foundation Grants
Nos. 2015M580965 and 2016T90028.
C. Zhang acknowledges NSFC under Grant Nos. 11504253 and 11734015.
A.~M.~O. kindly acknowledges support by Narodowe Centrum Nauki
(NCN, National Science Centre, Poland) under Project
No.~2016/23/B/ST3/00839.

\appendix*

\section{Diagonalization of the Hamiltonian}
\label{sec:DIAGONALIZATION}

We are considering a 1D quantum compass model with the three-site
(XZX$+$YZY) terms under staggered magnetic fields in Eq. (\ref{model}),
which can be rewritten as
\begin{eqnarray}
\cal{H}&=& \sum_{i=1}^{N/2}
\left(J_{1}^{}\sigma_{2i-1}^x \sigma_{2i}^x
+J_{2}^{}\sigma_{2i}^y \sigma_{2i+1}^y \right)\nonumber \\
&+&\sum_{i=1}^{N/2}\left(
h_1\,\hat{z}\cdot\vec{\sigma}_{2i-1}+h_2\,\hat{z}\cdot\vec{\sigma}_{2i}\right)
\nonumber \\
&+&J_3\sum_{i=1}^{N}
\left(\sigma^x_{i-1} \sigma^z_{i}\sigma^x_{i+1}
+\sigma^y_{i-1} \sigma^z_{i}\sigma^y_{i+1}\right).
\label{Hamiltonian9}
\end{eqnarray}
Here, $J_1$ and $J_2$ denote the coupling strength on odd and even
bonds, respectively. $h_1$ and $h_2$ are the transverse external
magnetic fields applied on odd and even sites, respectively.
Finally, $J_3$ is the strength of (XZX$+$YZY)-type three-site
exchange interactions.

First, we use a Jordan-Wigner transformation which maps explicitly a
pseudo-spin model to a free-fermion system whose properties can always
be computed efficiently as a function of system size \cite{EBarouch70}:
\begin{eqnarray}
\sigma_j^z&=&1-2c_j^\dagger c_j^{}, \nonumber\\
\sigma_j^x&=&e^{i\phi_j} \left(c_j^\dagger + c_j^{}\right), \nonumber\\
\sigma_j^y&=&ie^{i\phi_j}\left(c_j^\dagger - c_j^{}\right),
\end{eqnarray}
with $\phi_j$ being the phase accumulated by all earlier sites, i.e.,
$\phi_j=\pi \sum_{l<j}c_l^\dagger c_l^{}$. Consequently, we have a
simple bilinear form of Hamiltonian in terms of spinless fermions:
\begin{eqnarray}
\cal{H}\!&=&\!\!\sum_{i=1}^{N/2}\!\left\{
J_1\!\left[\left(c_{2i-1}^\dagger c_{2i}^{}\!-c_{2i-1}^{}c_{2i}^{\dagger}\right)\!
+\!\left(c_{2i-1}^\dagger c_{2i}^\dagger\!-c_{2i-1}^{}c_{2i}^{}\right)\right]\right.
\nonumber \\
&+&\left. J_2\left[\left(c_{2i}^\dagger c_{2i+1}^{}-c_{2i}^{}c_{2i+1}^{\dagger}\right)
-\left(c_{2i}^\dagger c_{2i+1}^\dagger-c_{2i}^{}c_{2i+1}^{}\right)\right]\right.
\nonumber \\
&+& \left. h_1\left(1- 2c_{2i-1}^{\dagger}c_{2i-1}^{}\right)
          +h_2\left(1- 2c_{2i}^{\dagger} c_{2i}^{}\right)\right.\nonumber \\
&+&\left. 2J_3\left(c_{2j-1}^{\dagger}c_{2j+1}^{}+c_{2j}^{\dagger}c_{2j+2}^{}
+{\rm H.c.}\right)\right\}.
\label{Hamiltonian2}
\end{eqnarray}
The fermion version of this model corresponds to a dimerised
$p$-wave superconductor, in which the electrons also generate
next-nearest neighbor hopping. Such a two-component 1D Fermi gas on
a lattice is realizable with current technology, for example on an
optical lattice by using a Fermi-Bose mixture in the strong-coupling
limit~\cite{Lewenstein04}.

Following the standard Jordan-Wigner transformation, we rewrite the
Hamiltonian in the momentum space by taking a discrete Fourier
transformation for plural spin sites with the periodic boundary
condition (PBC).
\begin{equation}
c_{2j-1}\!=\!\sqrt{\frac{2}{N}}\sum_{k}e^{-ik j}a_{k},\quad
c_{2j}\!=\!\sqrt{\frac{2}{N}}\sum_{k}e^{-ik j}b_{k},
\end{equation}
with discrete momenta as
\begin{equation}
k\!=\frac{2n\pi}{ N  }, \quad n= -\left(\frac{N}{2}-1\!\right),
-\left(\frac{N}{2}-3\!\right),\ldots,\left(\frac{N}{2} -1\!\right)\!.
\end{equation}
Next the discrete Fourier transformation for plural spin sites is
introduced for the PBC. The Hamiltonian takes the following form
which is suitable to apply the Bogoliubov transformation:
\begin{eqnarray}
{\cal H}&=& \sum_{k} \left[ T_k a_{k}^{\dagger}b_{k}^{}-T_k^*a_{k}^{}
b_{k}^{\dagger}+T_k a_{k}^{\dagger}b_{-k}^{\dagger}
-T_k^* a_{k}^{}b_{-k}^{} \right.  \nonumber \\
&+&\left. F_k\left(a_k^{\dagger}a_k^{}+b_k^{\dagger}b_k^{}\right)
+2\delta\left(a_k^{\dagger}a_k^{}-b_k^{\dagger}b_k^{}\right)
+Nh\right],\nonumber \\
\label{Hamiltonian6}
\end{eqnarray}
where $T_k=J_1+J_2 e^{ik}$ and $F_k=2J_3\cos k-2h$. After the Fourier
transformation, $\cal{H}$ is then transformed into a sum of commuting
Hamiltonians $\hat{H}_k$ describing a different $k$ mode each. Then we
write the Hamiltonian in the BdG form in terms of Nambu spinors:
\begin{eqnarray}
\cal{H} &=&  \frac{1}{2}\sum_{k}
\Upsilon_k^{\dagger}
\hat{H}_k
\Upsilon_k, \label{FT2}
\end{eqnarray}
where
\begin{eqnarray}
\hat{H}_k = \left(\begin{array}{cccc}
  F_k+2\delta &  T_k &  0   &  T_k  \\
   T_k^* &  F_k-2\delta & -T_{-k}    & 0     \\
    0 &  -T_{-k}^*    & -F_k-2\delta & -T_k\\
     T_k^*   & 0   &  -T_k^* & -F_k+2\delta
   \end{array}\right),
\label{APP:HamiltonianMatrix2}
\end{eqnarray}
and $\Upsilon_k^{\dagger}=
(a_{k}^{\dagger},b_k^{\dagger},a_{-k}^{},b_{-k}^{})$.
In momentum space, time reversal (TR) symmetry and particle-hole (PH)
symmetry of the BdG Hamiltonian $\hat{H}_k$ are implemented by
anti-unitary operators $T$ and $\mathbf{C}$. For spinless fermions TR
operator $T$ is simply a complex conjugation $\cal{K}$ and operator
$\mathbf{C}=\tau_x K$ as the PH transformation. The system Eq.
(\ref{APP:HamiltonianMatrix2}) belongs to topological class $D$ with
topological invariant $\mathbb{Z}_2$ in one dimension, which satisfies
$\hat{H}(-k){\cal C}=-{\cal C}\hat{H}(k)$. Here
${\cal C}=\tau^x\otimes\sigma^0\cal{K}$, where $\tau^x$ and $\sigma^0$
are the Pauli matrices acting on PH space and spin space, respectively.

The diagonalized form of $\hat{H}_k$ can be achieved by a
four-dimensional Bogoliubov transformation which connects the original
operators $\{a_k^{\dagger},b_k^{\dagger} ,a_{-k}^{},b_{-k}^{}\}$, with
two kind of quasiparticles, $\{\gamma_{k,1}^{\dagger},
\gamma_{k,2}^{\dagger},\gamma_{-k,1}^{},\gamma_{-k,2}^{}\}$,
as follows,
\begin{eqnarray}
\left(
\begin{array}{c}
\gamma_{k,1}^{\dagger} \\
\gamma_{k,2}^{ \dagger}\\
\gamma_{-k,1}^{}  \\
\gamma_{-k,2}^{}
\end{array}
\right)=\hat{U}_{k} \left(
\begin{array}{c}
a_k^{\dagger}  \\
b_k^{\dagger}   \\
a_{-k}^{}   \\
b_{-k}^{}
\end{array}%
\right). \label{eq:UT}
\end{eqnarray}
$\hat{H}_k$ is diagonalized by a
unitary transformation (\ref{eq:UT}),
\begin{eqnarray}
{\cal H}= \sum_{k}
\Upsilon_{k}^{\dagger} \hat{U}_{k}^{} \hat{U}_{k}^\dagger \hat{H}_k
\hat{U}_{k}^{} \hat{U}_{k}^\dagger\Upsilon_{k}^{}=
\sum_{k}\Upsilon_{k}^{\prime\dagger} D_{k}^{} \Upsilon_{k}^{\prime}. \quad
\end{eqnarray}
The obtained four eigenenergies $\{\varepsilon_{k,j}\}$ ($j=1,\cdots,4$)
\begin{eqnarray}
\varepsilon_{k,1 (2)}&=&\sqrt{\vert T_k \vert^2+ F_k^2}\pm
\sqrt{\vert T_k\vert^2 +4 \delta^2}, \nonumber\\
\varepsilon_{k,4 (3)}&=&-\sqrt{\vert T_k \vert^2 +F_k^2}\mp
\sqrt{\vert T_k\vert^2 +4 \delta^2}, \label{APP:spec}
\end{eqnarray}
in the diagonalized Hamiltonian matrix
$D_{k}^{}=\hat{U}_{k}^\dagger\hat{H}_k\hat{U}_{k}^{}$
are the excitations in the artificially enlarged PH space where the
positive (negative) ones denote the electron (hole) excitations. The
ground state corresponds to the state in which all hole modes are
occupied while the electron modes are vacant. The PH symmetry indicates
here that
$\gamma_{k,4}^{\dagger}=\gamma_{-k,1}^{}$ and
$\gamma_{k,3}^{\dagger}=\gamma_{-k,2}^{}$.
So the spectra consist of two branches of energies
$\varepsilon_{k,j}$ (with $j=1,2$), and
\begin{eqnarray}
\hat{H}_{k} &=&\frac{1}{2}\,\varepsilon_{k,1}\left(
\gamma_{k,1}^{\dagger}\gamma_{k,1}^{} -
\gamma_{-k,1}^{}\gamma_{-k,1}^{\dagger}\right) \nonumber \\
&+&\frac{1}{2}\,\varepsilon_{k,2}\left(
\gamma_{k,2}^{\dagger}\gamma_{k,2}^{} -
\gamma_{-k,2}^{}\gamma_{-k,2}^{\dagger}\right) \nonumber \\
&=& \sum_{j=1}^2\,\varepsilon_{k,j}
\left(\gamma_{k,j}^{\dagger}\gamma_{k,j}^{}-\frac12\right).
\label{exc}
\end{eqnarray}

Two-point correlation functions for the real Hamiltonian Eq.
(\ref{Hamiltonian9}) can be expressed as an expansion of
Pfaffians using the Wick theorem,
\begin{eqnarray}
\langle \sigma_i^x \sigma_j^x \rangle &=& \left \vert
\begin{array}{c c c c}
G_{-1} & G_{-2} & \cdot & G_{i-j} \\
G_{0} & G_{-1} & \cdot & G_{i-j+1} \\
\vdots & \vdots & \ddots &\vdots  \\
G_{j-i-2} & G_{j-i-3} & \cdot & G_{-1}
\end{array} \right \vert, \\
\langle \sigma_i^y \sigma_j^y \rangle &=& \left \vert
\begin{array}{c c c c}
G_{1} & G_{0} & \cdot & G_{i-j+2} \\
G_{2} & G_{1} & \cdot & G_{i-j+3} \\
\vdots & \vdots & \ddots &\vdots  \\
G_{j-i} & G_{j-i-1} & \cdot & G_{1}
\end{array} \right \vert, \\
\langle \sigma_i^z\sigma_j^z \rangle &=&
\langle\sigma_i^z\rangle \langle\sigma_j^z\rangle- G_{j-i} G_{i-j},
\end{eqnarray}
where $G_r=\langle(c_0^\dagger-c_0^{})(c_r^\dagger+c_r^{})\rangle$
and $r=j-i$ represents the distance between the two sites in units of
the lattice constant.

\end{document}